\documentclass[apl, showpacs, preprint, preprintnumbers,amsmath,amssymb,superscriptaddress,endfloats]{revtex4-1}
\usepackage[]{graphicx}      % Include figure files
\usepackage{bm}             	% bold math
\usepackage{times}			% Times font
\usepackage{tabularx}
\usepackage{color}
\newcommand{\IEF}{Institut d'Electronique Fondamentale, CNRS, UMR 8622, Orsay, France}
\newcommand{\UPS}{Univ. Paris-Sud, 91405 Orsay, France}
\newcommand{\Nancy}{Institut Jean Lamour, CNRS - Universite� de Lorraine, Boulevard des aiguillettes BP 70239,
F-54506 Vandoeuvre le`s Nancy, France}
\newcommand{\Singulus}{Singulus technology AG, Hanauer Landstrasse 103, 63796 Kahl am Main, Germany}

\begin{document}
%\title ANCIENNE VERSION {Measurements of magnetization in CoFeB films using dipolar repulsion of domain walls}
\title{Measurement of magnetization using domain compressibility in CoFeB films with perpendicular anisotropy}

\author{N. Vernier}
\email{nicolas.vernier@u-psud.fr}
\affiliation{\UPS}
\affiliation{\IEF}
\author{J.-P. Adam}
\author{S. Eimer}
\affiliation{\IEF}
\affiliation{\UPS}
\author{G. Agnus}
\affiliation{\UPS}
\affiliation{\IEF}
\author{T. Devolder}
\affiliation{\IEF}
\affiliation{\UPS}
\author{T. Hauet}
\affiliation{\Nancy}
\author{B. Ockert}
\affiliation{\Singulus}
\author{D. Ravelosona}
\affiliation{\IEF}
\affiliation{\UPS}

\date{\today}                                           
%%%%%%%%%%%%%%%%%%%%%%%%%%%%%%%%%%%%%%%%
%
%       Abstract
%
%%%%%%%%%%%%%%%%%%%%%%%%%%%%%%%%%%%%%%%%
\begin{abstract}
We present a method to map the saturation magnetization of soft ultrathin films with perpendicular anisotropy, and we illustrate it to assess the compositional dependence of the magnetization of CoFeB(1 nm)/MgO films. The method relies on the measurement of the dipolar repulsion of parallel domain walls that define a linear domain. The film magnetization is linked to the field compressibility of the domain. The method also yields the minimal distance between two walls before their merging, which sets a practical limit to the storage density in spintronic devices using domain walls as storage entities.
\end{abstract}

%\PACS{: 06.60.Mr, 07.55.Jg, 75.60.Ch, 75.70.Ak}

\maketitle

%%%%%%%%%%%%%%%%%%%%%%%%%%%%%%%%%%%%%%%%
%
%                Paper
%
%%%%%%%%%%%%%%%%%%%%%%%%%%%%%%%%%%%%%%%%

Recently there has been a renewed interest in spintronic devices relying on the motion of narrow domain walls in magnetic nanowires. This includes the use of domain walls as storage units  \cite{Parkin:Science:2008, Zhang:JAP:2012, Uhlir:JPhysD:2012} or as information vectors performing logic operations \cite{Allwood:Science:2005,Jaworowicz:Nanotechnology:2009}. Since they combine a high perpendicular anisotropy \cite{Ikeda:NatMat:2010} with a coercivity \cite{Devolder:APL:2013a} lower than the standard systems exhibiting Perpendicular Magnetic Anisotropy (PMA), ultrathin CoFeB/MgO films are a promising system to study the motion of narrow domain walls. Indeed, walls in CoFeB/MgO systems are mobile \cite{Devolder:APL:2013a} in fields as low as 0.1 mT, and their motion seems not to be influenced by pinning phenomena for fields above 1 mT. 

To fine tune the properties of such films, one can play with the Boron content \cite{Munakata:IEEETransMag:2005}, the Fe-to-Co composition \cite{Devolder:APL:2013a}, the degree of crystallization \cite{Lee:APL:2007, Pym:APL:2006}, or the degree of mixing at the interfaces \cite{Devolder:JAP:2013}. A key feature to compare the performance of these films is their saturation magnetization $M_s$ and its uniformity at the local scale. Conventional magnetometry methods like Superconduction Quantum Interferometer Devices (SQUID) or Vibrating Sample Magnetometers (VSM) can inherently only give the spatial average of the magnetization, and are prone to errors due to the parasitic magnetic signals coming from the substrate or the surface contamination. Methods based on torques or their field derivatives like FerroMagnetic Resonance (FMR) can not separate the contributions of PMA and demagnetization fields in the thin film geometry, and they only give a qualitative measurement of the sample inhomogeneity \cite{McMichael:JAP:1998}. 

Here, we present a flexible method to measure the magnetization of soft PMA films that is operative down to sizes a few $10 \times 10~\mu\textrm{m}^2$, and we illustrate it to assess the compositional dependence of $M_s$ of CoFeB/MgO films in both as-grown and annealed states. The method builds on M. Bauer's work \cite{Bauer:PRL:2005} and relies on the manipulation of two neighboring narrow domain walls \cite{Muratov:JAP:2008, Kubetzka:PRB:2003, Mascaro:APL:2010, Thomas:NatCom:2012}. The principle is the following. Dipolar interactions favor the (central) domain between the two walls, because the walls repel each other proportionally to the film magnetization. The walls' separation can be adjusted by an external field. The measurement of the field induced compressibility of the central domain by magneto-optical microscopy yields a calibration-free way of deriving the saturation magnetization and its spatial uniformity at the $10~\mu\textrm{m}$ scale.

We have studied the compositions Co$_{60}$Fe$_{20}$B$_{20}$, Co$_{40}$Fe$_{40}$B$_{20}$  and Co$_{20}$Fe$_{60}$B$_{20}$, with the layer of interest being part of substrate/Ta(5 nm)/CoFeB($t=1$ nm)/MgO(2 nm)/Ta(5 nm) multilayers. Each sample was studied before and after an annealing of two hours at 300$^0$C. 
	
\begin{figure}
\includegraphics[width=8.4cm]{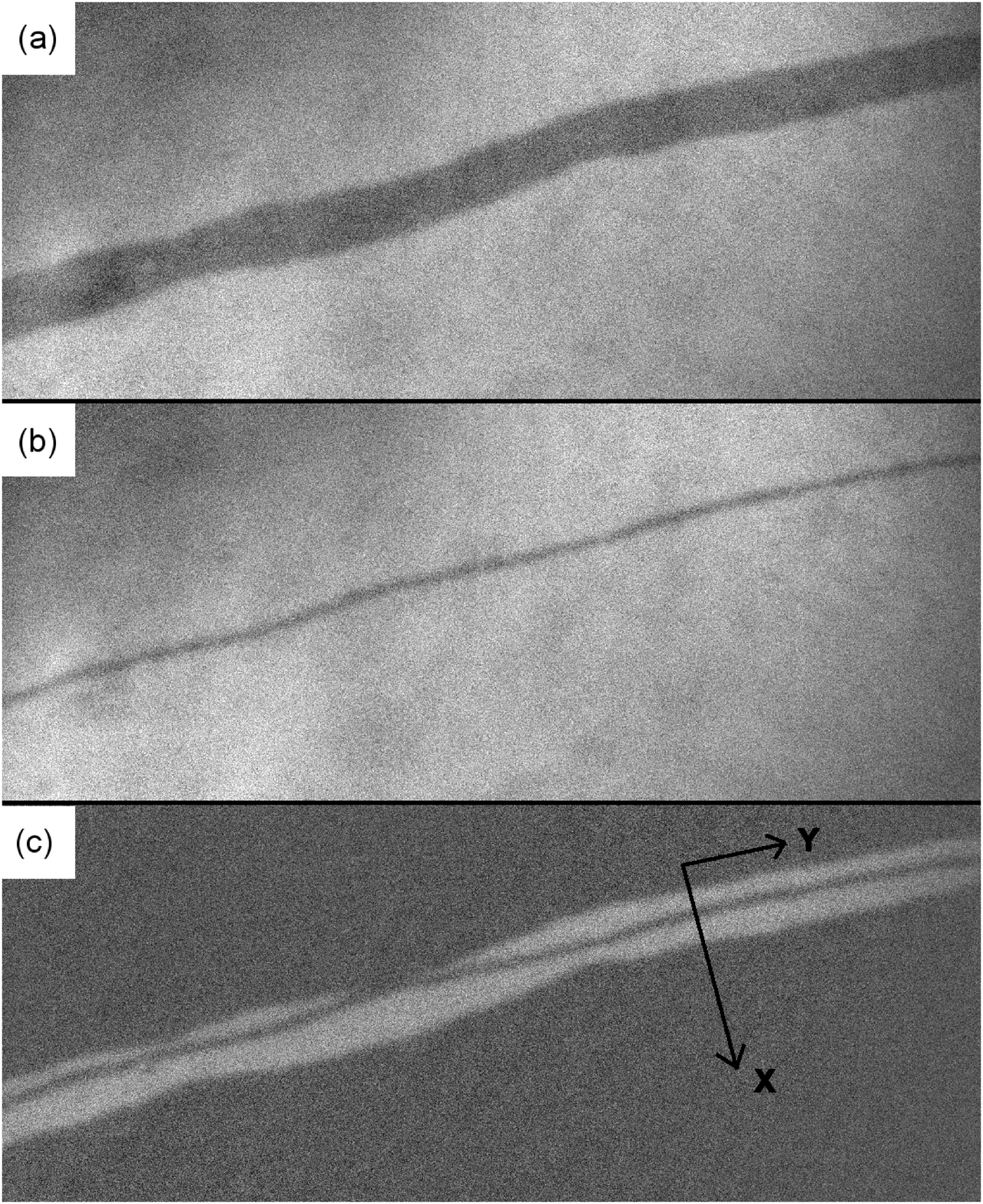}
\caption{Magneto-optical micrographs ($110\times45~\mu\textrm{m}^2$) of domain patterns in Co$_{20}$Fe$_{60}$B$_{20}$ (1 nm)/MgO films. (a) Pair of well separated ($d \approx 5~\mu$m) domain walls at remanence.  (b) Same pair of walls under a field of 0.59 mT. (c) Picture obtained by subtracting the two previous images.}
\label{images}
\end{figure}

The magnetic configurations were probed using a polar Kerr imaging setup, with a $\times 50$ magnification lens of numerical aperture 0.35. The nominal resolution according to Rayleigh criteria is $\delta \approx 0.8~\mu$m. In practice we shall look at \textit{linear} domain walls (Fig. \ref{images}), such that when there is a single domain wall its position $x_0$ can be identified with an accuracy much better than $\delta$, by simply fitting the optical profile with a step function of slope $\delta$. Experimentally, several step functions having a progressive transition appeared suitable, here, we have used $f(x) \propto \textrm{tan}^{-1}((x-x_0)/\delta)$. When several parallel domain walls are present, the finding of their positions is done using a deconvolution procedure which requires the exact knowledge of the contrast between the upward and downward magnetized states.  In practice, one thus needs to correct for the non-uniformity of the lightning and for the finite Faraday rotation of the objective lens. To cancel these artefacts, we have used the following experimental procedure.

%%
%	Figure 1
%%
%

\begin{figure}
\includegraphics[width=8.4cm]{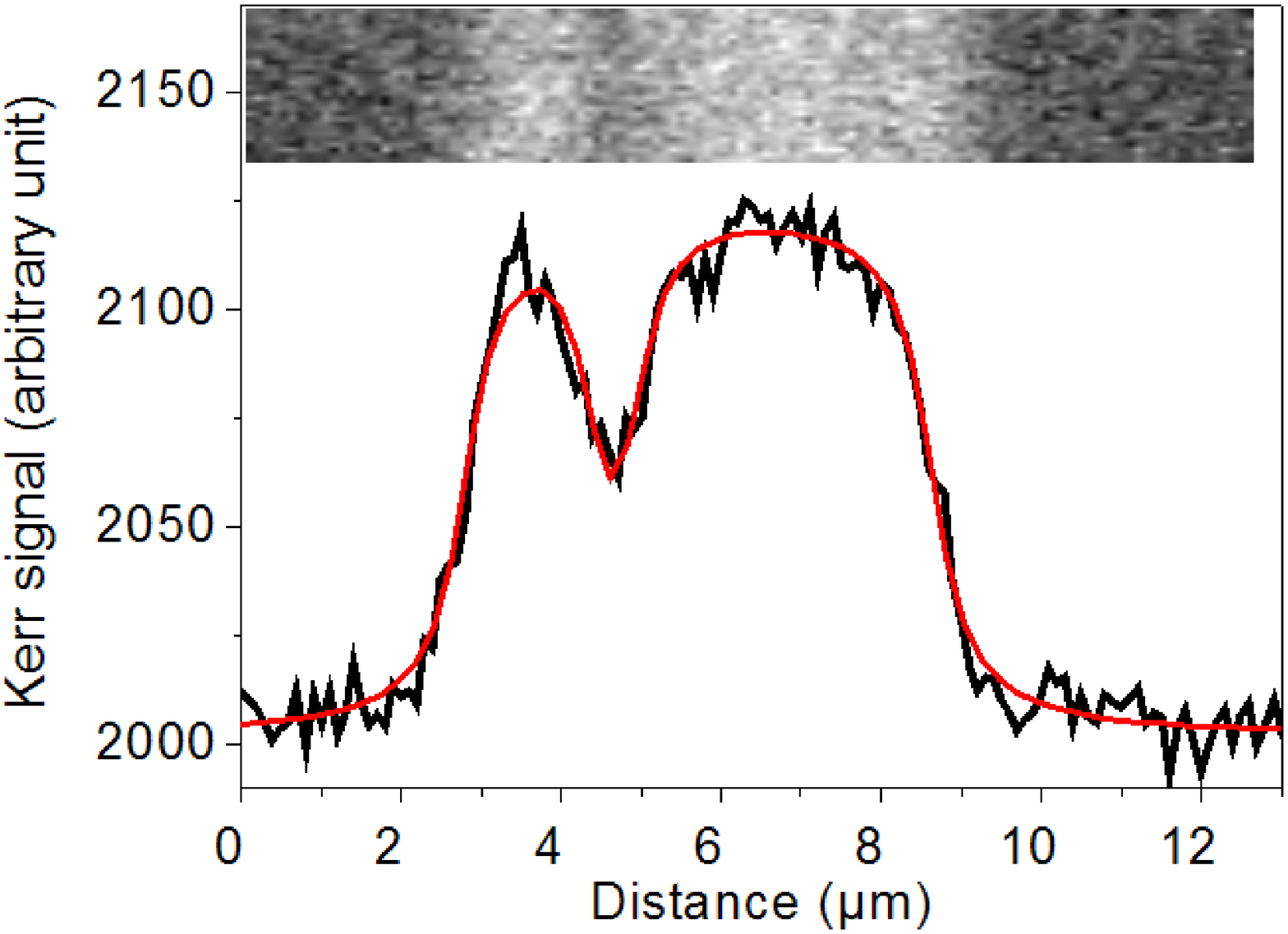}
\caption{(color online) Profile of the magneto-optical contrast obtained on the annealed  Co$_{20}$Fe$_{60}$B$_{20}$ (1 nm) sample, in a field of 0.59 mT. The walls positions found using Eq. \ref{riri} are $x_1=2.89$, $x_2=4.50$, $x_3=5.09$, and $x_4=8.89~\mu\textrm{m}$. The wall-to-wall separation is thus $d=590$ nm. Inset: magneto-optical image ($12.2\times3.1~\mu\textrm{m}^2$) used to get the contrast profile.}
\label{profile}
\end{figure}

 We first prepare the two parallel walls at positions $x_1$ and $x_4$ at a large distance from each other (i.e. $|x_4-x_1| >> \delta$).  This is done by almost saturating the sample, then enlarging the few remaining unreversed domains, and freezing them back in zero field (Fig.\ref{images}a).  Second, we apply an external field to compress the central domain (Fig.\ref{images}b). The walls are now positioned at $x_2$ and $x_3$. %If the wall separation $d$ is less than 1 $\mu$m, the contrast of the central domain is reduced (see Fig. \ref{images}b). 
 To estimate the new domain width $d=|x_3-x_2|$, we subtract Fig.\ref{images}a from Fig.\ref{images}b, getting Fig. \ref{images}c. %In this last image, the dark and bright parts give the grey levels for upward and downwards magnetization. 
 A stripe cut through $x$ (see Fig. \ref{images}c) yields a contrast profile (Fig. \ref{profile}) with plateaus accounting for the signals of a full reversal. The width of the stripe cut is chosen to mitigate the noise. In the example of Fig. \ref{profile}, the central domain is narrower than the optical resolution (i.e. $d < \delta$), such that the corresponding negative peak at $x \approx 5 ~\mu$m in the contrast profile does not reach the lower plateau. To get $d=|x_3-x_2|$, one fits the contrast profile (Fig. \ref{profile}) with a function 
 
 \begin{equation}
 c(x) = A_0+A_1 \sum_{i=1}^4 (-1)^i arctan\frac{(x_i-x)}{\delta}
 \label{riri}
 \end{equation}
 
\begin{figure}
\includegraphics[width=8.4cm]{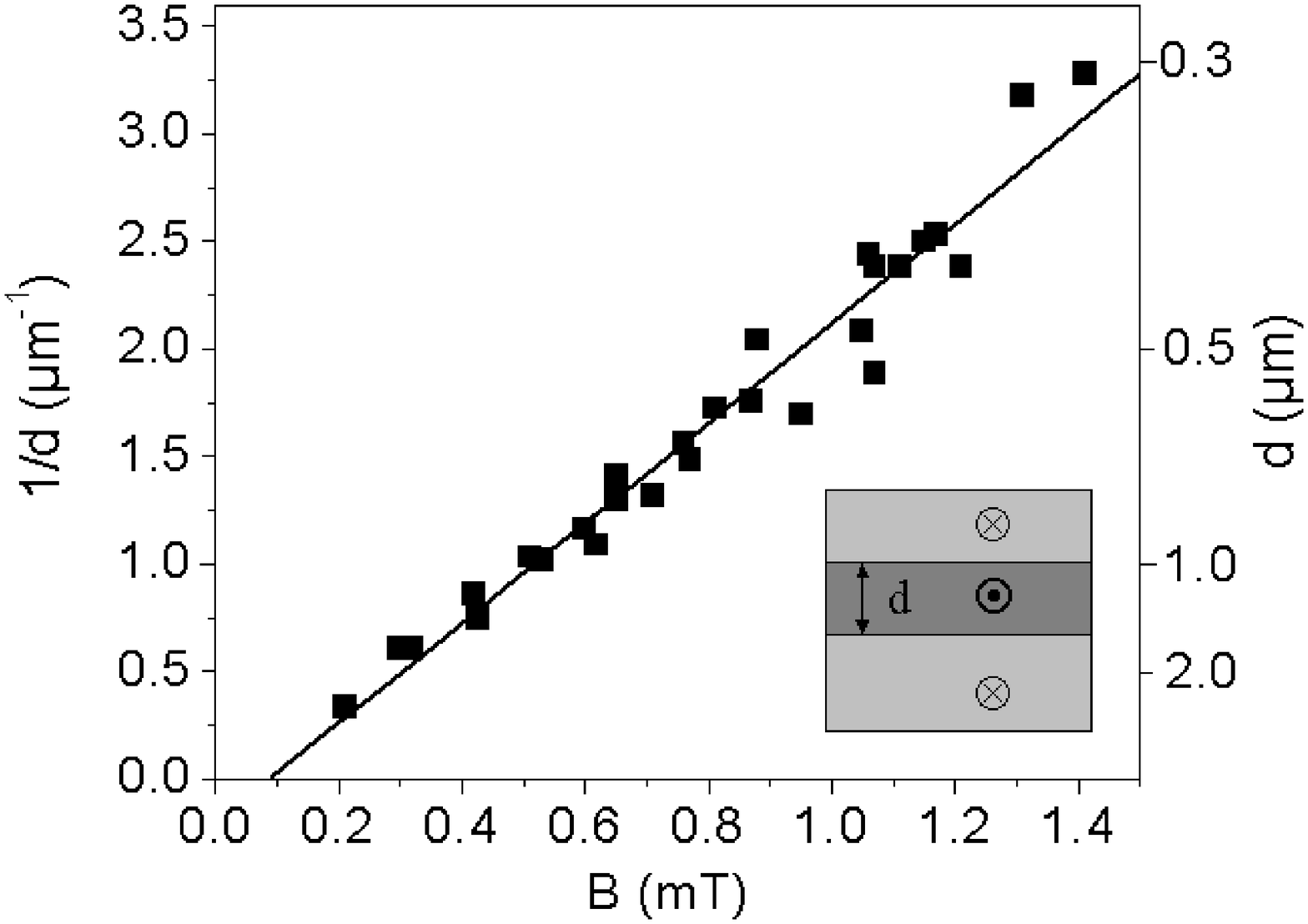}
\caption{Dependence of the wall to wall distance with the applied field for the as -grown Co$_{20}$Fe$_{60}$B$_{20}$ sample. The slope is the compressibility of the central wall, which measures the inverse magnetization. Inset: sketch of the domain structure.}
\label{regression}
\end{figure}

The adjustable parameters are the four wall positions $x_i$, the optical resolution $\delta$, the contrast scale $A_1$ and on offset $A_0$. We estimate that central domain size is known with an accuracy of $\pm$25 nm. This number was certified with specially designed samples consisting of thin aluminum wires on silicon with variable widths ranging from 100 to 1000 nm.
Finally, we emphasize that the measurement procedure is repeated as various places of the sample till we get a statistically reliable estimate of the dependance of $d$ with the applied field $H_{ext}$ (Fig. \ref{regression}). This minimizes the uncertainty associated to the wall roughness that is generally observed and results from pinning effects.

Let us now use the field dependance of the size of the central domain to get the film magnetization.
If the domain wall width $\Delta$ with is much smaller than the distance between the two walls, the wall-wall interaction is purely of dipolar origin and it is repulsive \cite{Muratov:JAP:2008, Wiebel:JAP:2006, Braun:PRB:1994, Bauer:PRL:2005}. Here we shall consider wall-to-wall distances greater than 300 nm (see Fig. \ref{regression}) in high PMA systems \cite{Devolder:APL:2013a}) where we expect $\Delta \leq 30$ nm, such that this condition is fulfilled. Under that approximation, the repulsive force is the analog of the Laplace force between two wires each carrying a charge current $I=2 t M_S$ and placed at separation $d$. On a given wall, the dipolar force per unit length is thus ${\mu_0 I^2}/({2 \pi d}) $. The film finite thickness term (see ref. \onlinecite{Braun:PRB:1994}) can be neglected in our case because our wall separation is substantially larger. An additional term exits in case. However in the presence of an external field there is an additional Zeeman pressure tending to compress the central domain. This force per unit length is $-2 \mu_0 M_S t H_{ext}$. In a defect free sample, these two forces would cancel each other when $t M_S = \pi d H_{ext}$. However in real films, a finite propagation field $\mu_0 H_p \approx 0.1$ mT is needed to overcome pinning effects and to induce domain wall motion. As a result there is an hysteresis in $d$ as a function of the sweeping direction of the external field. Assuming the distance to be measured with a field compressing the central domain, the wall to wall distance $d$ is :
\begin{equation}
	d^{-1} = \pi  \frac {H_{ext}-H_p}  {t M_S}							
\label{zozo}
\end{equation}

Linear fits of through Eq. \ref{zozo} yield $M_s$, as examplified in Fig. \ref{regression}.

%%
%	Figure 2
%%
%

Table \ref{bilan} gathers the magnetizations independently obtained using either our present method or conventional magnetometry on larger samples (at least $2\times 2$ mm$^2$), on the various compositions of CoFeB. The values are given before and after annealing except for the as-grown Co$_{60}$Fe$_{20}$B$_{20}$ sample because it showed in-plane easy axis. A satisfactory agreement is found between the magnetization values deduced from SQUID, AGFM, and domain compressibility. However the die to die dispersion of the $M_S t$ values make us suspect the existence of composition and/or thickness fluctuations across the wafer, especially for the Co-rich compositions. These possible structural variations may exacerbate the inhomogeneity of the magnetization because of the proximity to the face-centered-cubic to hexagonal-compact phase boundary \cite{Schreiber:SSC:1995} in the FeCo binary alloy phase diagram. In all cases, annealing slightly increases the magnetization, confirming previous reports \cite{Lam:JM:2013, Yamanouchi:IEEETML:2011, Bilzer:JAP:2006}.

The compositions leading to the highest magnetizations are Co$_{20}$Fe$_{60}$B$_{20}$ and  Co$_{40}$Fe$_{40}$B$_{20}$. Position of ternary alloys on the Slater-Pauling curve is not obvious \cite{Galanakis:JPhysD:2006, Fecher:JAP:2006}, but it seems that boron has little influence on the magnetic properties apart from a dilution effect \cite{Munakata:IEEETransMag:2005}. From the Slater-Pauling curve, a broad maximum of magnetization for a ratio of cobalt of around 28\% is expected (corresponds to 35\% for a Boron-free CoFe alloy), which is compatible with our findings (Table \ref{bilan}).

%%
%	Figure 3
%%
%

During these experiment, we have been able to measure two additional interesting quantities. The first quantity is the magnetic field needed to merge the two neighboring domain walls and let the central domain disappear abruptly. We emphasize that although two different configurations are expected depending on the winding directions of each wall, a unique critical field was measured: statistical measurements indicated that this critical field is a reproducible metric, reported in Table \ref{bilan}. Above these applied fields, the number of domain walls changes inside a given sample: the data integrity in domain wall based memories \cite{Parkin:Science:2008} is then lost, which gives the working boundaries of such devices if based on soft PMA systems like ours. Besides, applying Eq. \ref{zozo} at this critical destruction field yields the second interesting quantity: the minimal stable wall-to-wall distance, found between 180 and 500 nm, depending on sample (Table \ref{bilan}). The measurement of this minimum wall separation $d_{min}$ is interesting from both applied and fundamental points of view. Indeed $d_{min}$ could be indicative of the effective profile of 180$^0$ domain walls since the disappearance of the central domain may just occur when the two walls are about to start overlapping. Also, this minimal wall-to-wall distance $d_{min}$ sets a practical limit to the storage density in racetrack memory applications \cite{Parkin:Science:2008}.

In summary, we have presented a calibration-free method to measure the local magnetization in ultrathin magnetic film having perpendicular anisotropy. The local character of the method could be used to a great advantage to measure the magnetization on patterned samples, for which the sensitivity of conventional magnetometry methods is not sufficient. We have illustrated our method by studying the composition dependence of the magnetization of CoFeB ultrathin films. In addition, our method yields the minimal achievable stable distance between two domain walls in such soft films, which sets the storage density limit in memory paradigms based on domain walls.
The authors wish to thank Jean-Pierre Jamet and Jacques Ferr\'{e} for useful discussions. This work was  supported by the European Communities FP7 program through contract MAGWIRE number 257707.

\begin{table*}
  \centering
  \begin{tabular}{|c|c|c|c|c|c|l}
  \hline
  Sample 	& a-Co$_{20}$Fe$_{60}$B$_{20}$ &  c-Co$_{20}$Fe$_{60}$B$_{20}$ & a-Co$_{40}$Fe$_{40}$B$_{20}$ & c-Co$_{40}$Fe$_{40}$B$_{20}$ & c-Co$_{60}$Fe$_{20}$B$_{20}$ \\ \hline  
 $\mu_0 M_S$ (T) (SQUID, 1$^{st}$ and 2$^{nd}$ meas.)  & 1.38 & (1.5)1.41 & 1.26 &  1.38 & (0.9)1.1  \\  \hline 
 $\mu_0 M_S$ (T) (AGFM) & 1.3 &1.1 &1.15 &1.3 & 0.8  \\  \hline 
 $\mu_0 M_S$ (T) (average of the above values) &1.34 &1.34 &1.20 &1.34 &0.93 \\  \hline 
$\mu_0 M_S$ (T) (present method) &1.35 &1.5 &1.25 &1.65 &0.825 \\  \hline \hline
Critical destruction field (mT) &1.3 &2.2 &1.7 &2.9 &0.6 \\ \hline
{Minimal wall to wall distance} $d_{min}$ & 355 nm & 220 nm & 260nm & 185 nm & 470 nm \\ \hline

    \end{tabular}
  \caption{synthesis of the results obtained on the different samples. a- stands for as-grown (amorphous) samples. c- stands for annealed (crystalline) samples. The sample a-Co$_{60}$Fe$_{20}$B$_{20}$ is not presented here because it was an in-plane anisotropy sample. }
  \label{bilan}
\end{table*}

\newpage
\printtables
\newpage
\printfigures

\end{document}